# On the charge transfer between single-walled carbon nanotubes and graphene

Rahul Rao,[a] Neal Pierce, Archi Dasgupta[#]

*Honda Research Institute USA, Columbus, Ohio 43212, USA*

**Abstract**

It is important to understand the electronic interaction between single-walled carbon nanotubes (SWNTs) and graphene in order to use them efficiently in multifunctional hybrid devices. Here we deposited SWNT bundles on graphene-covered copper and $SiO_2$ substrates by chemical vapor deposition and investigated the charge transfer between them by Raman spectroscopy. Our results revealed that, on both copper and $SiO_2$ substrates, graphene donates electrons to the SWNTs, resulting in p-type doped graphene and n-type doped SWNTs.

_______________

a) Electronic Mail: rrao@honda-ri.com
[#] On leave from Penn State University, State College, PA, USA

SWNTs and graphene are unique nanostructured forms of carbon that possess a variety of fascinating electrical, mechanical and chemical properties.[1] It is therefore of great interest to use them in hybrid multifunctional architectures where the unique properties of both materials are utilized synergistically. Indeed there have been several recent reports in the literature on hybrids between nanotubes and graphene in a variety of different configurations,[2-4] and that have exhibited superior performances in applications such as energy storage and optics.[5-7] The nanotube/graphene hybrid devices rely on electronic interactions between the constituents, and it is important to understand these interactions at a fundamental level. Graphene is a semi-metal, whereas SWNTs can be either semiconducting or metallic.[1] Moreover, it is known that substrates and the environment play important roles in regulating the charge densities of both graphene and SWNTs; for example, adsorption of oxygen typically causes doping by positive charge carriers (holes).[8,9]

Raman spectroscopy is a powerful tool that has been used extensively to study charge transfer in SWNTs and in graphene.[10-19] Adsorption of electron donating and accepting molecules cause shifts in the Raman peak frequencies, which are proportional to the amount of carriers injected.[10] While charge transfer in SWNTs and graphene due to molecular dopants has been studied separately, there is very little work on the electronic interaction between SWNTs and graphene. One recent study[18] reported the interaction between an individual metallic SWNT and graphene, and found the graphene to be doped n-type by the SWNT. In their study the SWNT was first deposited on a $SiO_2$ substrate, followed by transfer of chemical vapor deposition (CVD) grown graphene on top of the SWNT. For applications, however, one must consider macroscopic architectures of SWNTs and graphene involving both semiconducting and metallic tubes. Furthermore, the role of the substrate (especially conductive substrates) cannot be neglected when considering charge transfer between SWNTs and graphene.

To this end, here we investigated the electronic interaction between SWNT films and graphene on metal (copper) substrates, an architecture that is attractive for applications such as energy storage. We thus studied the fundamental charge transfer processes that occur due to the matching of Fermi levels between SWNTs, graphene and the Cu substrate in assembled hybrid architectures. Our samples were made by the deposition of SWNT bundles grown via CVD directly on to graphene-coated Cu substrates (obtained from Graphene Laboratories Inc.). The electronic interaction between the SWNTs and graphene was studied by Raman spectroscopy. In addition, the graphene was also transferred from Cu to $SiO_2$ substrates, followed by SWNT deposition in order to compare the effect of an insulating substrate on the charge densities in SWNTs and graphene.

It is important to note is that it is not straightforward to study charge transfer between SWNTs and graphene via Raman spectroscopy. Since both graphene and SWNTs consist of $sp^2$-bonded carbon, their Raman peaks overlap, making it difficult to distinguish between their peak frequencies. To circumvent this issue, we used [13]C isotope-labeled graphene (hereafter called [13]C-graphene) on Cu foils. The Raman peaks of [13]C-graphene are red-shifted from non isotope-labeled graphene by a factor equaling

√(12/13), making it possible to unambiguously identify the SWNT and graphene peaks.[2] We were thus able to accurately analyze peak shifts and thereby study the electronic interaction between SWNTs and $^{13}C$-graphene on Cu and $SiO_2$. For comparison, the control samples were as-prepared $^{13}C$-graphene on Cu, and SWNTs deposited on plain Cu foils.

The SWNTs were deposited on the $^{13}C$-graphene by floating catalyst CVD. The $^{13}C$-graphene-coated Cu and $SiO_2$ substrates were loaded downstream at the exhaust end of a 1" diameter quartz tube. The substrates were placed vertically upright so that they faced the flow of gases through the furnace. A syringe pump was connected to the inlet end of the quartz tube for delivery of the precursor. Mixtures of ferrocene (1 wt.%) and ethanol were injected at rates of 12-15 ml/hr into the hot zone of the tube furnace once it reached the growth temperature of 1000 ºC. The precursor evaporated at the inlet of the furnace and the iron and carbon species were carried through the furnace with a mixture of argon (500 sccm) and hydrogen (200 sccm). In the floating catalyst process, the SWNTs grow from iron particles in-flight and are deposited on substrates that are placed downstream. Since the substrates were well outside the hot zone of the furnace, their temperature only went up to ~40 ºC during the growth. The growth time was kept between 5 and 15 minutes so that the deposited SWNT film consisted of sparse individual SWNTs or small bundles. Reducing the amount of the deposited SWNTs allowed us to easily measure the Raman spectra from the graphene underneath. Following SWNT deposition, the samples were analyzed by scanning electron microscopy (SEM) and Raman spectroscopy with multiple excitation wavelengths (633, 532 nm). The laser power was kept low (<2 mW) for the Raman measurements in order to avoid the effects of heating.

Examples of Raman spectra collected with the 633nm excitation from the [13]C-graphene before (top trace) and after (bottom trace) SWNT deposition are shown in Fig. 1. The Raman spectrum from graphene typically exhibits two intense peaks, labeled the *G* and *G'* bands.[20] The *G* band corresponds to in-plane lattice vibrations of the carbon atoms in the graphene lattice. The *G'* band is a dispersive second order peak corresponding to scattering of two transverse optic (TO) phonons, and is typically much more intense than the *G* band in graphene. The *G* band typically appears at ~1585 cm$^{-1}$ in graphene on Cu; however it is red-shifted in [13]C-graphene by $\sqrt{(12/13)}$ due to the [13]C isotope. As can be seen in Fig. 1, the *G* band in the [13]C-graphene appears at ~1520 cm$^{-1}$. The *G'* band undergoes a similar redshift and appears at ~2530 cm$^{-1}$ (with the 633 nm excitation).

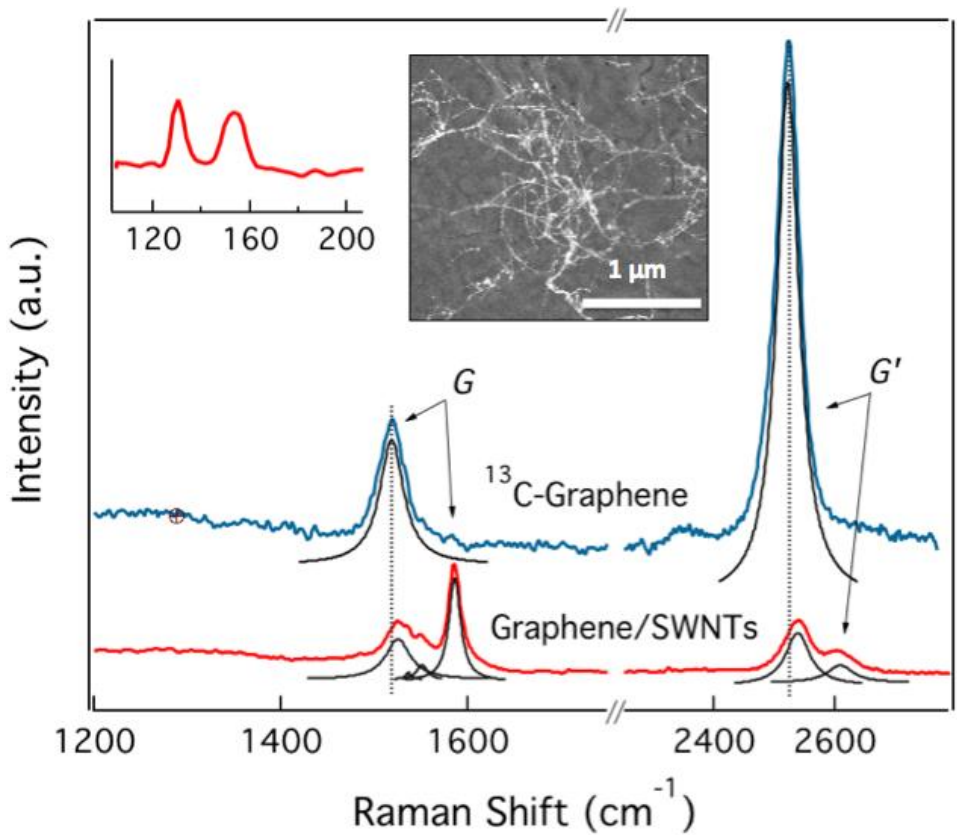

FIG. 1. Typical Raman spectra collected with 633 nm excitation from the [13]C-graphene on Cu before (top trace) and after (bottom trace) SWNT deposition. The *G* and *G'* bands have been fitted with Lorentzian peaks. The upper left inset shows the low frequency RBM peaks from the SWNTs on the [13]C-graphene. The upper right inset is an SEM image showing low-density SWNTs on [13]C-graphene.

The *G* band in a SWNT is more complicated than graphene; it corresponds to multiple vibrational modes tangential to the nanotube axis, with the most intense peak (called *G⁺*) appearing at ~1590 cm[-1].[20] Moreover, the Raman spectrum of the SWNT exhibits unique peaks in the low frequency region, called radial breathing modes (RBMs), which correspond to lattice vibrations in the radial direction. The RBM peak frequencies are inversely proportional to the nanotube diameters. The bottom trace in Fig. 1 shows an example of a Raman spectrum collected after SWNT deposition on the $^{13}$C-graphene. Both the *G* and *G'* peaks from the graphene and SWNTs are clearly distinguishable. Also shown in Fig. 1 (upper inset) are the low frequency RBMs from the SWNTs. At least two peaks can be observed at 130 and 153 cm$^{-1}$, indicating that the SWNT diameters range between 1-1.5 nm.[20] Similar Raman spectra were obtained from the SWNTs deposited on the $^{13}$C-graphene-covered $SiO_2$ substrates. An SEM image of the SWNTs on the $^{13}$C-graphene is shown in the upper right inset in Fig. 1. As mentioned above, the deposition times were kept short so that the SWNT film did not become too thick to prevent the measurement of the underlying graphene Raman peaks. The SWNTs observed in Fig. 1 appear to be individual or in least small bundles, along with some catalyst particles that inevitably decorate the SWNTs during deposition. The $G^+$ band from bundled SWNTs ($10 - 20$ cm$^{-1}$) is typically broader than for individual SWNTs (<10 cm$^{-1}$).[20] Furthermore, the linewidths of RBMs of individual SWNTs are very narrow (1-3 cm$^{-1}$).[21] In our spectra the average $G^+$ band linewidth is 15 cm$^{-1}$ and the RBM linewidths range from 5-10 cm$^{-1}$. The broadened peaks in addition to the observation of multiple RBMs supports the presence of bundles in the SWNTs. In addition to the *G* and G' bands, a third peak called the *D* band typically appears at ~1330 cm$^{-1}$ (with 633 nm excitation) in both the SWNT and graphene Raman spectrum due to disorder.[20] As can be seen in Fig. 1, the spectrum in the *D* band region exhibits negligible intensity after SWNT deposition, indicating that the SWNT deposition process did not cause any damage to the $^{13}$C-graphene.

The dotted lines in Fig. 1 clearly show that the Raman peaks of graphene under the SWNTs are blueshifted compared to their frequencies before SWNT deposition. These shifts in frequencies are more clearly illustrated in Fig. 2a, which plots the graphene $G'$ peak frequencies against the $G$ peak frequencies from several different samples. The data in Fig. 2a are a comparison of the Raman peaks from the [13]C-graphene on Cu, before and after SWNT deposition (labeled Graphene/SWNTs in Fig. 2a). In general both the graphene $G$ and $G'$ peaks from graphene are blueshifted after SWNT deposition, indicating hole doping in the graphene.[14] The average $G$ band frequency is blueshifted by 3 cm$^{-1}$, while the average $G'$ band frequency is blueshifted by 30 cm$^{-1}$ due to charge transfer from the SWNTs. It should be noted that the graphene is doped with holes even before SWNT deposition. Pristine un-doped [13]C-graphene exhibits a $G$ band at ~1520 cm$^{-1}$ (calculated with the assumption that the pristine [12]C-graphene $G$ band is at ~1580 cm$^{-1}$), which is 3 cm$^{-1}$ lower than the average $G$ band frequency observed in the [13]C-graphene on the Cu foil (1523 cm$^{-1}$). The dotted lines in Fig. 2a show the expected $G$ and $G'$ band frequencies for pristine (un-doped) [13]C-graphene. Graphene is known to be hole-doped due to the oxidized Cu substrate following CVD growth,[22] confirming the present observations. The data in Fig. 2a therefore suggests that the [13]C-graphene becomes hole doped due to the interaction with SWNTs. In other words, graphene donates its electrons to the SWNTs.

The Raman peaks from the SWNTs exhibit an equal and opposite behavior compared to the graphene peaks after SWNT deposition on the [13]C-graphene. Fig. 2b shows the $G$ and $G'$ peak frequencies collected from several samples. As mentioned above, the control samples were SWNT films deposited on plain Cu foils. Unlike the graphene Raman peaks which blueshift after SWNT deposition (Fig. 2a), the Raman peaks from the SWNTs redshift due to interaction with graphene (labeled SWNT/Graphene in Fig. 2b). The average $G$ band frequency in the SWNT/graphene is redshifted by 3 cm$^{-1}$ compared to the corresponding average value from SWNTs on plain Cu. The $G'$ band frequency is blueshifted by an average of 23 cm$^{-1}$. Similar to Fig. 2a, the dotted lines in Fig. 2b denote the frequencies of undoped SWNTs on Cu. Previously it has been shown that doping of SWNTs by electron donor species causes a redshift of the $G$ band frequency. [12,13,15] The observed redshift of the $G$ band in SWNTs/graphene can therefore be attributed to transfer of electrons from the [13]C-graphene to the SWNTs. The n-type doping of the SWNTs is also concomitant with the p-type doping observed in the graphene (Fig. 2).

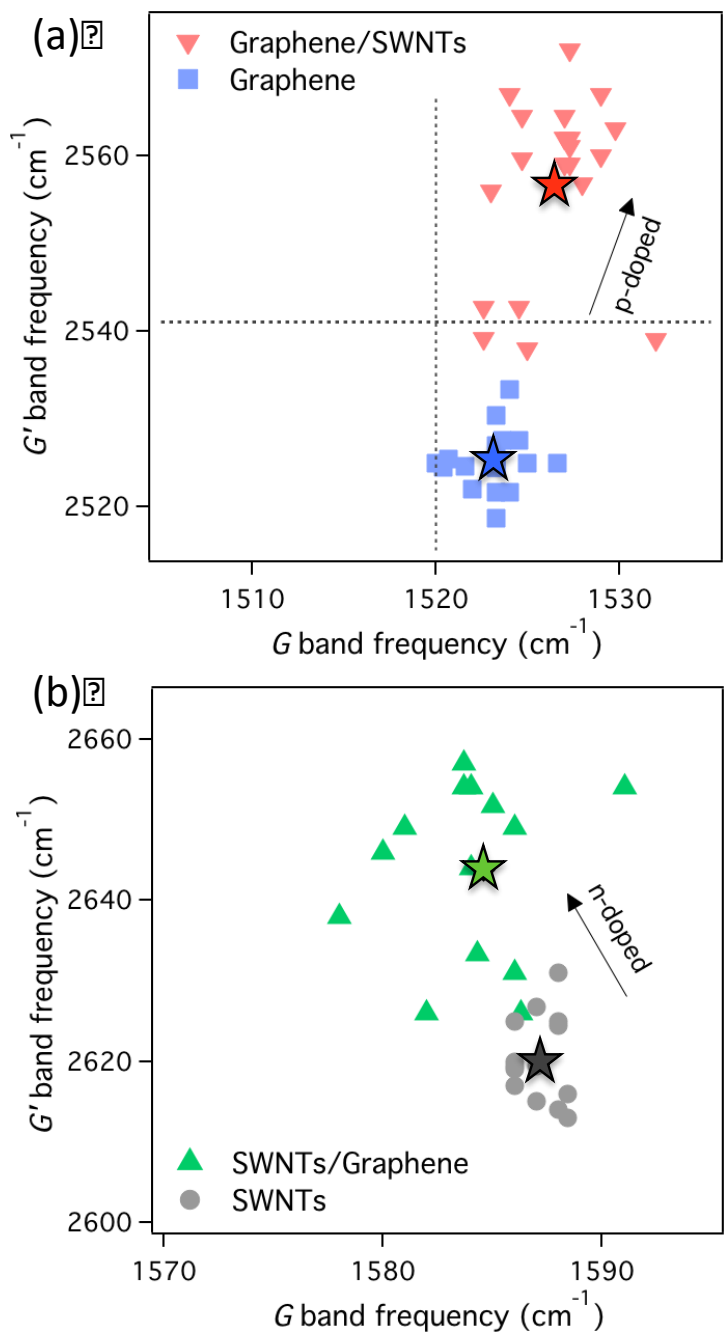

FIG. 2. (a) Plot of the *G′* band frequencies against the *G* band frequencies from the (a) ¹³C-graphene with and without SWNTs on Cu substrates, and (b) from the pristine SWNTs and SWNTs on ¹³C-graphene coated Cu. The ★ data points indicate the average values in the data sets. The dotted lines in (a) indicate the frequencies of the pristine (un-doped) ¹³C-graphene.

We also compared the Raman spectra from SWNTs deposited on $SiO_2$ substrates and graphene-coated $SiO_2$. The graphene was transferred from Cu to $SiO_2$ using the following procedure - A PMMA solution (Microchem 495 PMMA A2) was spin-coated on the graphene/copper foils at 4000 rpm for 60 s and then placed on a ~180 °C hot plate for 1 min. The graphene on the backside of the foil was removed by etching in a nitric acid solution (1 $HNO_3$: 3 $H_2O$) for 3 min. The copper was then removed with a 0.1 M ammonium persulfate solution. After rinsing the graphene film with distilled water, it was placed on an $SiO_2$ substrate. After air-drying, the graphene-on-$SiO_2$ was heated at 180 °C for 1 hour to minimize wrinkling of the graphene film. Finally, the PMMA was removed with an acetone bath.

The Raman spectra from the $SiO_2$ substrates exhibited shifts in the *G* and *G'* band frequencies similar to the Cu substrates, with some obvious differences. Figs. 3a and 3b show the *G* and *G'* band frequencies from the [13]C-graphene and SWNTs on $SiO_2$. Comparing the peak frequencies from the bare graphene (without SWNTs) on Cu (Fig. 2a) and on $SiO_2$ (Fig. 3a), it is apparent that even before SWNT deposition the doping levels in the graphene are different between the two substrates. Recall that the *G* band frequency in undoped [13]C-graphene lies at ~1520 $cm^{-1}$ (dotted vertical line in Fig. 3a). The *G* band frequencies in the graphene on $SiO_2$ vary from 1510 – 1525 $cm^{-1}$, indicating that the doping in graphene on $SiO_2$ varies from n-type to p-type depending on the spot/sample measured. The different doping levels are probably due to the transfer process (described above) from Cu to $SiO_2$, which involves PMMA and other organic solvents that are possibly only partially removed at the end of the process, leaving behind pockets of doped graphene. However, similar to the case of the Cu substrates, the Raman peaks of the $^{13}$C-graphene under the SWNTs are blueshifted. The average *G* and *G'* band frequencies in the graphene/SWNTs are blueshifted by 10 cm$^{-1}$ and 30 cm$^{-1}$, respectively (Fig. 3a). Thus the graphene is doped with excess holes from the SWNTs. The Raman peaks of the SWNTs redshift concomitantly, although the shift of the *G'* band is not as severe as on Cu. As can be seen in Fig. 3b, the G band from the SWNTs is redshifted by an average of 5 cm$^{-1}$, while the *G'* band is blueshifted by ~5 cm$^{-1}$.

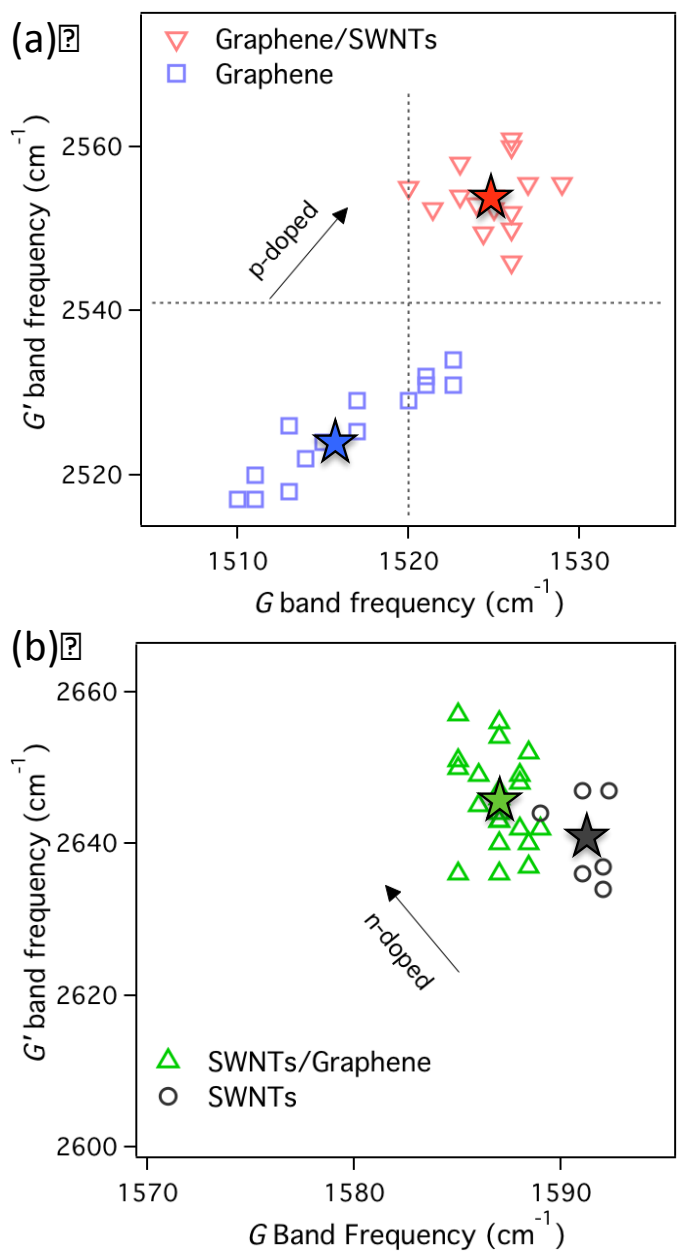

FIG. 3. (a) Plot of the $G'$ band frequencies against the $G$ band frequencies from the (a) $^{13}$C-graphene with and without SWNTs on SiO$_2$ substrates, and (b) from the pristine SWNTs and SWNTs on $^{13}$C-graphene coated SiO$_2$. The ★data points indicate the average values in the data sets. The dotted lines in (a) indicate the frequencies of the pristine (un-doped) $^{13}$C-graphene.

In spite of the differences in frequencies of the Raman peaks of the graphene and SWNTs on Cu and $SiO_2$, the direction of shifts of the graphene and SWNT peaks on $SiO_2$ is similar to those on Cu. The Raman analysis indicates p-type and n-type doping in the $^{13}C$-graphene and SWNTs, respectively. We also observe similar shifts in peak frequencies with a different laser excitation (532 nm, not shown here). Our results are in contrast to the previous Raman measurements performed on a graphene-covered individual metallic SWNT.[18] In their study, Paulus *et al.* found the graphene to be n-doped due to the charge transfer from the metallic SWNT. The graphene *G* band was redshifted by 10 $cm^{-1}$, from which they deduced an electron density of 1.12 x $10^{13}$ $cm^{-2}$ due to electron doping from the SWNT.

In our case the *G* band of the $^{13}C$-graphene is blueshifted by an average of 3 $cm^{-1}$ (maximum 9 $cm^{-1}$). Comparing our peak shift with data obtained on gated graphene devices,[14] we estimate the hole concentration in the graphene to be ~0.5 x $10^{13}$ $cm^{-2}$ (with a maximum of 1 x $10^{13}$ $cm^{-2}$). It is more challenging to determine the charge concentration in SWNTs based on shifts in the Raman spectra. The *G* band in SWNTs comprises of several peaks that correspond to different types of vibrations tangential to the axis of the SWNT. The two most intense peaks are labeled $G^-$ and $G^+$ for the lower and higher frequency peaks, respectively. The various changes in the Raman spectra due to doping include differences in the intensity ratios between the $G^-$ and $G^+$ peaks, broadening/stiffening of the peaks, and frequency shifts.[14,16,26] Moreover, the as-prepared SWNT bundles contain a mixture of semiconducting and metallic SWNTs,[10,15] which have different charges concentrations and charge transfer characteristics with graphene. Nevertheless, an average redshift of the SWNT *G* band frequency on Cu (3 cm$^{-1}$) and SiO$_2$ (5 cm$^{-1}$) indicates electron transfer to the SWNTs from the [13]C-graphene.

The p-type and n-type doping of the graphene and SWNTs make sense if one considers their work functions. The work function of graphene has been estimated to be ~4.6 eV[23] while that of SWNTs varies depending on the whether the SWNT is individual (semiconducting or metallic) or in bundles. The peaks in the Raman from bundled SWNTs are typically broader than peaks from individual SWNTs. The In our samples the SWNTs are either individual or in small bundles (Fig. 1). We can therefore use the work function determined for bundles, which is ~4.8 eV.[24] Since the work function is the difference in energy between the vacuum level and the Fermi level, a smaller work function implies a higher lying Fermi level and electrons will flow from this material to the one with the higher work function (lower lying Fermi level). Thus we expect electrons to flow from the graphene to the SWNTs where they contact each other. Moreover, the work functions of Cu and CuO are 4.65 and 5.3 eV,[25] respectively, which are both larger than that of graphene, implying that electrons flow from the graphene to the Cu. This reasoning is in accord with the experimental observation of p-type doping in graphene on Cu prior to SWNT deposition. As mentioned above, the present results contradict the findings of Paulus *et al*., where they found n-doping of graphene by the SWNT,[18] rather than p-doping. In their case the interaction of the graphene was with a single metallic SWNT, whose work function was estimated to be 4.5 eV, i.e. lower than that of graphene. The lower work function of the individual SWNT caused the electron doping of graphene.

The differences observed in the electronic interaction between an individual metallic SWNT (Ref. 18) and bundles (present results) highlight the uniqueness of the SWNT and graphene electronic structures, as well as the importance of studying the electronic interactions between graphene and SWNTs in different architectures. With our unique synthesis method, we were able to directly deposit SWNTs on graphene-coated Cu substrates, thereby simulating a hybrid SWNT-graphene device on a metal electrode. Our method can also be easily extended towards the deposition of SWNTs on other 2D layered materials, for example $MoS_2$, whose hybrid with SWNT films has recently been demonstrated as a gate-tuneable p-n junction.[26]

**Acknowledgements**

AD acknowledges financial support from the Honda Research Institute, USA Inc.